\documentclass[review,onefignum,onetabnum]{siamart171218}

\usepackage{amsfonts}
\usepackage{graphicx}
\usepackage{epstopdf}
\usepackage{booktabs}
\usepackage{algpseudocode}
\usepackage{xtab}
\usepackage{lipsum}

\usepackage{enumitem}
\usepackage{algorithm}

\usepackage{multirow}
\renewcommand{\vec}[1]{\mathbf{#1}}
\newcommand{\Jvec}{\vec{J}}
\newcommand{\nvec}{\vec{n}}
\newcommand{\R}{\mathbb{R}}

\newcommand{\Ichk}{I_{\text{ch},k}}
\newcommand{\Ich}{I_\text{ch}}

\newcommand{\jump}[1]{[#1]}
\newcommand{\jumpm}[1]{[#1]_m}
\newcommand{\avg}[1]{\{#1\}}

\newcommand{\mk}[1]{\textcolor{black}{#1}}

\DeclareMathOperator{\Div}{\nabla \cdot}
\DeclareMathOperator{\Grad}{\nabla}

\ifpdf
  \DeclareGraphicsExtensions{.eps,.pdf,.png,.jpg}
\else
  \DeclareGraphicsExtensions{.eps}
\fi

\newsiamremark{remark}{Remark}
\newsiamremark{hypothesis}{Hypothesis}
\crefname{hypothesis}{Hypothesis}{Hypotheses}
\crefname{subsection}{section}{sections}
\Crefname{subsection}{Section}{Sections}
\newsiamthm{claim}{Claim}

\headers{A DG solver for ionic electrodiffusion}{Ellingsrud et al.}

\title{A splitting, discontinuous Galerkin solver for the cell-by-cell electroneutral Nernst-Planck framework}

\author{ Ada J.~Ellingsrud\thanks{Department for Numerical Analysis and
Scientific Computing, Simula Research Laboratory, Oslo, Norway
(\email{ada@simula.no}, \email{benedp@simula.no}, \email{miroslav@simula.no})} \and Pietro
Benedusi\footnotemark[1] \and Miroslav
Kuchta\footnotemark[1]
}

\usepackage{amsopn}

\makeatletter
\newcommand*{\addFileDependency}[1]{
  \typeout{(#1)}
  \@addtofilelist{#1}
  \IfFileExists{#1}{}{\typeout{No file #1.}}
}
\makeatother

\ifpdf
\hypersetup{
  pdftitle={KNP-EMI-DG},
  pdfauthor={Ellingsrud and Kuchta}
}
\fi

\begin{document}

\maketitle


\begin{abstract}
  Mathematical models for excitable tissue with explicit representation of individual cells are
  highly detailed and can, unlike classical homogenized models, represent complex cellular geometries
  and local membrane variations. However, these cell-based models are challenging to approximate
  numerically, partly due to their mixed-dimensional nature with unknowns both in the bulk and at
  the lower-dimensional cellular membranes. 
  We here develop and evaluate a novel solution strategy for the cell-based KNP-EMI model describing ionic electrodiffusion
  in and between intra- and extracellular compartments with explicit representation of individual cells.
  The strategy is based on operator splitting,  a multiplier-free formulation of the coupled dynamics across sub-regions,
  and a discontinuous Galerkin discretization. In addition to desirable theoretical properties, such as local mass conservation,
  the scheme is practical as it requires no specialized functionality in the finite element assembly and 
  order optimal solvers for the resulting linear systems can be realized with black-box algebraic multigrid
  preconditioners. Numerical investigations show that the proposed solution strategy
  is accurate, robust with respect to discretization parameters, and that the
  parallel scalability of the solver is close to optimal -- both for idealized and
  realistic two and three dimensional geometries. 

\end{abstract}




\section{Introduction} Despite the fundamental role of movements of molecules
and ions in and between cellular compartments for brain
function~\cite{rasmussen2020interstitial}, most computational models for
excitable tissue assume constant ion concentrations~\cite{Koch1999, Rall1977,
sterratt2023principles}.  Although these models have provided valuable insight
into how neurons function and communicate, they fail to describe vital
processes related to ionic signalling and brain
homeostasis~\cite{nicholson1978calcium, Chen2000}, and pathologies involving
substantial changes in the extracellular ion composition such as epilepsy and
spreading depression~\cite{Dietzel1989, Somjen2001, Sykova2008}. The emerging
KNP-EMI framework~\cite{ellingsrud2020finite, mori2009numerical} is a system of partial differential equations (PDEs) describing the
coupling of ion concentration dynamics and electrical properties in excitable
tissue with explicit representation of the cells, allowing for morphologically
detailed descriptions of the neuropil. In contrast to classical homogenized
models, the highly detailed KNP-EMI framework enables modelling of single
cells and small collections of cells, uneven distributions of membrane mechanisms,
and the role of the cellular morphology in tissue dynamics.



Electrodiffusive transport of each ion species in the intracelluar space (ICS) and extracellular space (ECS) in the KNP-EMI framework is described by a Nernst Planck (NP) equation. To determine the electrical potential in the bulk, the NP equations
are coupled with an electroneutrality assumption stating that there is no
charge separation anywhere in solution~\cite{pods2017comparison,
  dickinson2011electroneutrality}. Consequently, the medium is electroneutral everywhere circumventing the need for
resolving charge relaxation processes at the nanoscale. Further, the cellular membranes are explicitly represented as
lower-dimensional interfaces separating the intra- and extracellular sub-regions. Coupling conditions at the
membrane interface relate the dynamics in the different sub-regions. The resulting problem is
mixed-dimensional, containing unknowns in both the intra- and extracellular domains (e.g.~electrical potentials)
and on the interface/cell membranes (e.g.~transmembrane currents). We note that the interface
is typically a manifold of co-dimension 1 with respect to the ICS/ECS.
Further, the system is
non-linear, coupled, and stiff: the fast electrical dynamics and the slow ion diffusion are associated with
vastly different time scales. This suggests split solution approaches 
  where the different dynamics are loosely coupled.
Nevertheless, the numerical strategies for approximating the system must be chosen with care.

Previously, the KNP-EMI problem has been discretized with a mortar finite element method (FEM),
  leading to a saddle-point problem where the intra- and extracellular potentials and
  concentrations are coupled together via Lagrange multipliers on common interfaces~\cite{ellingsrud2020finite}.
  The resulting formulation is challenging for implementation as it necessitates, e.g.~support for different finite element spaces on domains/meshes with different topological dimension (ECS, ICS and the membranes)
  and their coupling.
  In addition, efficient solvers for the resulting linear systems require specialized multigrid
  methods \cite{wohlmuth2005calv} or preconditioners \cite{Kuchta2021}.
  Other discretization approaches for the KNP-EMI model, or its sub-problem
  the cell-by-cell (EMI) equations~\cite{krassowska1994response, ying2007hybrid, agudelo2012numerical},
  have been considered, including,
  finite volume schemes \cite{mori2009numerical, xylouris2010three},
  boundary element methods \cite{rosilho2024boundary},
  CutFEM finite element methods \cite{berre2023cut} or
  finite differences \cite{tveito2017cell}. Efficient solvers for the
  different formulations of the EMI model have been developed
  e.g.~in~\cite{huynh2023convergence, budisa2023algebraic, benedusi2024EMI, rosilho2024boundary}.


\mk{
We here present a novel solution strategy for the KNP-EMI problem
where the Lagrange multipliers are eliminated and only the bulk variables are
explicitly solved for. As all the problem unknowns are then posed over domains
of same topological dimension, we refer to this formulation as \emph{single-dimensional}.
The single-dimensional formulation may be discretized with
conforming elements, e.g.~continuous Lagrange elements in~\cite{benedusi2024scalable}, but several factors motivate
us to apply discontinuous Galerkin (DG) schemes instead.
In particular, standard DG functionality (e.g.~facet integrals) is sufficient to implement to
DG schemes for KNP-EMI. This is in contrast to conforming discretization where specialized mixed-mesh coupling
across ECS and ICS meshes is required.
Moreover, DG methods inherently allow for adaptable refinement
and have desirable properties, such as local mass conservation or numerical stability
in convection-dominated regime~\cite{riviere2008discontinuous}.
%
%
Our approach further includes an operator splitting technique decoupling the concentrations
from the electrical potential reducing the system to two simpler sub-problems: (i) an EMI problem,
and (ii) a series of advection diffusion (KNP) problems, which can leverage dedicated, recently developed solvers, e.g.~\cite{roy2023scalable, tveito2021modeling, miroDG}.
}

A key feature of the KNP-EMI models is the ability to handle complex and
realistic geometries, often resulting in large-scale systems when discretized.
\mk{
  In turn, efficient and scalable solvers are required to utilize the modeling
  potential of the framework in practical applications. As will become apparent,
  our splitting approach and DG discretization address these issues as
  the resulting linear systems are amenable to block-box algebraic multigrid (AMG)
  preconditioners. Here, the order optimal solvers can thus be realized as
  AMG-preconditioned Krylov methods.
}


Through numerical investigations, we demonstrate that our proposed numerical scheme is accurate and yields optimal convergence rates in space and time. In idealised 2D and 3D geometries, we further show that the proposed preconditioners are robust with respect to numerical parameters.
Experiments show that the parallel scalability of our solution scheme is close to optimal, enabling large-scale simulations on HPC clusters. Finally, we assess quantities of interest, such as conduction
velocity, ECS potentials and concentration shifts in a physiological relevant scenario where we simulate neuronal activity in a morphological realistic 3D geometry representing a pyramidal neuron in the cortex.

\section{Mathematical framework} 
\label{sec:KNP-EMI}
We here present the coupled, time-dependent,
non-linear, mixed-dimensional KNP-EMI equations describing ionic electrodiffusion in a geometrically explicit setting. For further details on the derivation of the equations, see e.g.~\cite{mori2009numerical, ellingsrud2020finite}.

\subsection{Governing equations}\label{sec:formulation}
We consider $N-1$ domains $\Omega_{i^n} \subset \R^d$ ($d = 1, 2, 3$) for $n =
1, \dots, N-1$ representing disjoint intracellular regions (physiological
cells) and an extracellular region $\Omega_e$, and let the complete domain be
$\Omega = \Omega_{i^1} \cup \dots \cup \Omega_{i^N} \cup \Omega_e$ with
boundary $\partial \Omega$. We denote the cell membrane associated with cell
$i^n$, i.e.~the boundary of the physiological cell $\Omega_{i^n}$, by
$\Gamma_n$. We assume that $\Gamma_n \cap \Gamma_m = \emptyset$ for all $n \not
= m$ and that $\Gamma_n \cap \partial \Omega = \emptyset$. \mk{Below we will denote the
restriction of functions $u: \Omega \rightarrow \R$, $f_l: \Omega \rightarrow \R$
to $\Omega_r$ by $u_{r}$ and $f_{l, r}$ respectively.}
For notational simplicity and
clarity, we present in the following the mathematical model for one intracellular region
$\Omega_{i^1} = \Omega_{i}$ with membrane $\Gamma$.

We consider a set of ion species $K$. For each ion species $k \in K$, we aim to find the
\emph{ion concentrations} $c_k: \Omega \times (0,T] \rightarrow \R $ , the
\emph{electrical potentials} $\phi: \Omega \times (0,T] \rightarrow \R$ and the
\emph{total transmembrane ion current} $I_M: \Gamma \times (0,T] \rightarrow \R$ such that:
\begin{align}
  \label{eq:concentration:org} 
    \frac{\partial c_k}{\partial t} + \Div \Jvec_k &= 0
    &\qquad \text{in } \Omega, \\
  \label{eq:phi:org}
    F \sum_{k \in K} z_k \Div \Jvec_k &= 0 &\qquad \text{in } \Omega,  \\
  \label{eq:mem:potentialrel}
    \phi_M &= \phi_i - \phi_e &\qquad \text{ on } \Gamma,   \\
  \label{eq:membrane:conservation}
    I_M\equiv - F \sum_{k \in K} z_k \Jvec_{k,e} \cdot \nvec_e
   &=  F \sum_{k \in K} z_k \Jvec_{k,i} \cdot \nvec_i  &\qquad \text{ on } \Gamma,\\
  \label{eq:mem:conc:i}
    \Jvec_{k,i} \cdot \nvec_i &= 
    \frac{\Ichk + \alpha_{k,i} (I_M- \Ich) }{F z_k} &\qquad \text{ on } \Gamma, \\
  \label{eq:mem:conc:e}
    - \Jvec_{k,e} \cdot \nvec_e &= 
    \frac{\Ichk + \alpha_{k,e} (I_M- \Ich) }{F z_k} &\qquad \text{ on }
    \Gamma,  \\
  \label{eq:mem:potential}
    \frac{\partial \phi_M}{ \partial t} &= \frac{1}{C_M}(I_M- I_\text{ch}) &\qquad
    \text{ on } \Gamma,
    \end{align}
where each regional ion flux density $\Jvec_k: \Omega \times (0,T] \rightarrow \R^d$ can be
expressed by a Nernst-Planck equation as follows:
\begin{equation}
  \label{eq:fluxJ:org}
  \Jvec_k = - D_k \nabla c_k - c_k z_k D_k \psi \nabla \phi,
  \qquad \text{in } \Omega
\end{equation}
stating that ionic movement is driven by both diffusion due to ionic gradients (first term) and drift in the electrical field (second term). Note that the drift term may be interpreted as an advective term where $z_k D_k \psi \nabla \phi$ drives the advection. The constant $\psi = F / (R T)$ combines the effective diffusion coefficient
$D_k$, Faraday's constant $F$, the absolute temperature $T$, and the gas
constant $R$.  
Further the membrane capacitance, the valence and the diffusion
coefficient for ion species $k$ are denoted by $C_M$, $z_k$ and $D_k$,
respectively. In the coupling conditions \eqref{eq:mem:conc:i}--\eqref{eq:mem:conc:e}, we assume that the ion specific capacitive current is some fraction $\alpha_k: \Gamma \times (0,T] \rightarrow \R$ of the total capacitive current where
\begin{equation}
    \label{eq:alpha}
    \alpha_k = \frac{D_k z_k^2 c_k}{\sum_{l \in K}D_l z_l^2 c_l}
    \qquad \text{ on } \Gamma.
\end{equation}
The ion specific channel currents are denoted by $\Ichk: \Gamma \times (0, T] \rightarrow \R$ and will depend on the \textit{membrane potential} $\phi_M: \Gamma \times (0, T] \rightarrow \R$
and typically be of the form
\begin{align}
\label{eq:Ichk}
\Ichk = g_k(\phi_M - E_k), \qquad E_k = \frac{RT}{z_k F} \ln{ \frac{c_{k,e}}{c_{k,i}}}  \qquad \text{ on } \Gamma,
\end{align}
where $g_k$ and $E_k$ denotes respectively the channel conductance and the Nernst potential for ion species $k$. The total ion current $\Ich: \Gamma \times (0, T] \rightarrow \R$ is given by $\Ich = \sum_k \Ichk$. In many physiological scenarios, the ion specific currents $\Ichk$ will further depend on \textit{gating variables} $\mathbf{s}: \Gamma \times (0, T] \rightarrow \R^{S}$, where $S$ is the number of gating variables, governed by ordinary differential equations (ODEs) of the following form:
\begin{align} 
    \label{eq:ODE:pot}
    \frac{\partial \phi_M}{\partial t} = \frac{1}{C_M}(I_M- \Ich) \qquad \text{ on } \Gamma, \\
    \label{eq:ODE:gate} 
    \frac{\partial \mathbf{s}}{\partial t} = F(\phi_M, \mathbf{s}) \qquad \text{ on } \Gamma.
\end{align}
We will refer to membrane models depending on variables governed by ODEs as active.

\subsection{Formulation with fewer concentrations}
\label{sec:alt:formulation}
The equation for the electrical potential~\eqref{eq:phi:org} is derived assuming
electroneutrality (see e.g.~\cite{mori2009numerical, roy2023scalable} for further details).
The electroneutrality assumption states that there is no charge separation in the bulk, and that the bulk of the tissue is electroneutral:
\begin{align}
    \label{eq:electroneutrality}
    0 = \sum_{k=1}^{K} z_k c_k \qquad \text{in } \Omega.
\end{align}
Using~\eqref{eq:electroneutrality}, we can express one of the ion concentrations as:
\begin{align}
    \label{eq:elim:cm}
    c_m = - \frac{1}{z_m} \sum_{k=1}^{m-1} z_k c_k \qquad \text{in } \Omega.
\end{align}
By inserting ~\eqref{eq:elim:cm} into~\eqref{eq:phi:org} and~\eqref{eq:membrane:conservation}, we can eliminate $c_m$ from the system. Let $\tilde{K}$ be $K \setminus\{m\}$. 
The modified system reads: for each $k \in \tilde{K}$, find the 
\emph{ion concentrations} $c_k: \Omega \times (0,T] \rightarrow \R$, the
\emph{electrical potentials} $\phi: \Omega \times (0,T]  \rightarrow \R$ and the
\emph{total transmembrane ion current} $I_M: \Gamma \times (0,T] \rightarrow \R$ such that~\eqref{eq:concentration:org}--\eqref{eq:mem:potential} hold. The eliminated concentration $c_m$ can be recovered via~\eqref{eq:elim:cm}. 


\subsection{Boundary and initial conditions}
\label{sec:bcs}
The system must be closed by appropriate initial and boundary conditions. We assume that initial conditions are given for the ion concentrations $c_k = c_k(\vec{x},t)$ for $k \in K$ and for the membrane potential $\phi_M = \phi_M(\vec{x},t)$:
\begin{align}
\begin{split}
        c_k(\vec{x}, 0) &= c_k^{0} =
        \begin{cases}
            c_{k,i}^{0}  &\quad \text{if } \vec{x} \in \Omega_i\\
            c_{k,e}^{0}  &\quad \text{if } \vec{x} \in \Omega_e,
        \end{cases}
    \end{split} \\
    \phi_M(\vec{x}, 0) &= \phi_M^{0} \qquad \qquad \quad \quad \vec{x} \in \Gamma,
\end{align}
where the initial concentrations $c_k^{0}$ satisfy the electroneutrality assumption, i.e.:
\begin{align}
    \label{eq:electroneutrality:ic}
    0 = \sum_{k=1}^{K} z_k c_k^{0} \qquad \text{in } \Omega.
\end{align}
Finally, we state that no ions can leave or enter the system by imposing the following Neumann boundary condition for $k \in K$:  
\begin{align}
    \label{eq:boundary:cond}
    \Jvec_{k}(\vec{x},t) \cdot \nvec = 0 \qquad  \vec{x} \in \partial \Omega.
\end{align}
\mk{
  We note with \eqref{eq:boundary:cond} the electrical
  potential in \eqref{eq:concentration:org}-\eqref{eq:mem:potential}
  is determined only up to a constant.
}

\section{Numerical scheme}
\label{sec:numerical:scheme}
We here consider a single-dimensional formulation of the KNP-EMI equations
where the total transmembrane ion current variable $I_M$ is eliminated.
Our approach is further based on a decoupling of the Nernst-Planck
equations~\eqref{eq:concentration:org} for ionic transport and
equation~\eqref{eq:phi:org} governing the electrical potential, as well 
an operator splitting scheme to
decouple the ODEs from the PDEs. For spatial discretization we will finally
employ the DG finite element method. 

\subsection{Temporal discretization and splitting scheme}
\label{sec:time:discretization}
We start by discretizing the system in time. For each time step $t^{n}$, we assume that the concentrations $c_k^{n-1}$ and the
membrane potential $\phi_M^{n-1}$ are known for $t^{n-1}$. The time derivatives in~\eqref{eq:concentration:org} and~\eqref{eq:mem:potential} are approximated respectively by
\begin{align}
    \label{eq:time:deriv}
\frac{\partial c_k}{\partial t} \approx
    \frac{c_k^n - c_k^{n-1}}{\Delta t} \quad \text{and} \quad 
    \frac{\partial \phi_M}{\partial t} \approx
    \frac{\phi_M^n - \phi_M^{n-1}}{\Delta t}
\end{align}
with step size $\Delta t$. 
To solve~\eqref{eq:concentration:org} and \eqref{eq:phi:org} at time $t^{n}$, we
apply the following two step splitting scheme where $c_k^n$ and $\phi^n$ denote respectively the ion concentrations and electrical potential at $t^n$:

{
\setlist[itemize]{leftmargin=20mm}

\begin{itemize}
  \item[\textbf{Step I:}] Find $\phi^n$ such that:
\begin{align}
    \label{eq:phi:split}
    F \sum_{k \in K} z_k \Div {\Jvec_k^n} &= 0, \\
    \label{eq:phi:split:J}
    {\Jvec_k^n} &= - D_k \nabla c_k^{n-1} - z_k D_k \psi c_k^{n-1} \nabla \phi^n.
\end{align}


\item[\textbf{Step II:}]
Find $c_k^n$ such that:
\begin{align}
    \label{eq:con:split}
    {\Delta t}^{-1}c_k^n + \Div \Jvec_k^n &= {\Delta t}^{-1}c_k^{n-1}, \\
    \label{eq:con:split:J}
    \Jvec_k^n &= - D_k \nabla c_k^n - z_k D_k \psi c_k^n \nabla \phi^n
\end{align}
for $k \in \tilde{K}$. Update $c_k^{n-1} = c_k^n$ and 
$\phi_M^{n-1} = \phi^n_i - \phi^n_e$.
\end{itemize}
}
\mk{
\noindent In both steps, the Neumann boundary conditions \eqref{eq:boundary:cond} are
assumed. Consequently, the first problem determining the electrical potential
$\phi^n$ is singular with constants in the nullspace. Moreover, its solvability
requires that the right-hand-side in \eqref{eq:phi:split}-\eqref{eq:phi:split:J}
satisfies certain compatibility condition. This point will be made more specific
when we discuss linear solvers in \Cref{sec:solver}.
}

Next, we eliminate the unknown total transmembrane ion current $I_M$ via the temporal discretization of~\eqref{eq:mem:potential} to obtain a single-dimensional form. The time derivative in~\eqref{eq:mem:potential} is approximated by~\eqref{eq:time:deriv} and the coefficients $\alpha_k$, the Nernst potentials $E_k$, and the ionic currents $\Ichk$ are treated explicitly:
\begin{align*}
    \alpha_k & \approx \alpha_k^{n-1} = \frac{D_k z_k^2 c_k^{n-1}}{\sum_{l \in K}D_l z_l^2 c_l^{n-1}}, \qquad
    E_k \approx E_k^{n-1} = \frac{RT}{z_k F} \ln{ \frac{c_{k,e}^{n-1}}{c_{k,i}^{n-1}}}, \\
    \Ichk &\approx \Ichk^{n-1} = \Ichk(\phi_M^{n-1}, c_k^{n-1}).
\end{align*}
For notational simplicity, we omit the temporal superscript and denote $\alpha^{n-1}_k$ by $\alpha_k$, $E_k^{n-1}$ by $E_k$, and $\Ichk^{n-1}$ by $\Ichk$ below. Similar convention will be applied also to other variables. By inserting~\eqref{eq:mem:potentialrel} into the discrete counterpart of~\eqref{eq:mem:potential} we
obtain the following expression for the total transmembrane ion current $I_M$:
\begin{align}
    \label{eq:IM:new}
    I_M= \frac{C_M}{\Delta t}(\phi_i^n - \phi_e^n - f) \qquad \text{ on } \Gamma,
\end{align}
where 
\begin{align}
\label{eq:f}
f = f^{n-1} = \phi_M^{n-1} - \frac{\Delta t}{C_M} \Ich.
\end{align}
\mk{
  We observe that \eqref{eq:IM:new} represents a Robin-type interface condition for
  $\phi^n_i$ and $\phi^n_e$. Specifically, 
}
by inserting~\eqref{eq:IM:new} into~\eqref{eq:mem:conc:i} and~\eqref{eq:mem:conc:e}
we get the following new expression for the ion specific fluxes across the cellular membrane interface:
\begin{align}
\label{eq:mem:conc:i:new}
    \Jvec_{k,i} \cdot \nvec_i =
    \frac{\alpha_{k,i} C_M}{F z_k \Delta t}(\phi_i^n - \phi_e^n - g_{k,i}) \qquad \text{ on } \Gamma, \\
  \label{eq:mem:conc:e:new}
    - \Jvec_{k,e} \cdot \nvec_e =
    \frac{\alpha_{k,e} C_M}{F z_k \Delta t}(\phi_i^n - \phi_e^n - g_{k,e}) \qquad \text{ on } \Gamma,
\end{align}
where $g_{k,i}$ and $g_{k,e}$ denote the restrictions of $g_k$ to respectively $\Omega_i$ and $\Omega_e$ with
\begin{equation}
    \label{eq:g}
    g_k = g_k^{n-1} = \phi_M^{n-1} - \frac{\Delta t}{C_M \alpha_{k}}.
\end{equation}
Note that in the case of an active membrane model (governed by ODEs) the expressions for $f$ and $g_k$ will change, as we shall see later in \Cref{sec:split:active}.

\subsection{Spatial discretization}

\begin{figure}
  \centering
  \includegraphics[width=1.0\textwidth]{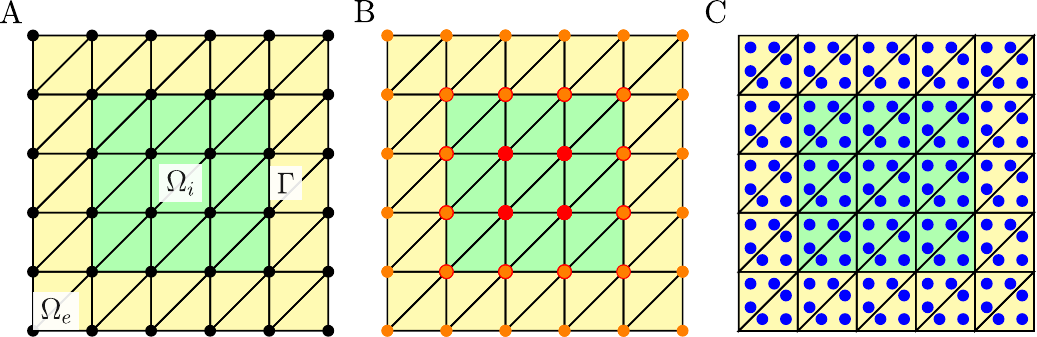}
  \vspace{-20pt}
  \caption{
    Finite element discretization of the KNP-EMI model by 
    continuous linear and discontinuous linear Lagrange elements.
    Setup with single intracellular domain $\Omega_i$ (in green) is considered.
    Triangulation conforms to the interface $\Gamma$. Mesh vertices are
    shown with black circle markers (\textbf{A}). Intra-/extracellular variables are represented in \emph{separate}
    $H^1$-conforming FE spaces over different meshes of $\Omega_i$ and $\Omega_e$.
    The coupling requires specialized implementation as on $\Gamma$ the degrees of
    freedom (shown with red and orange markers) interact (\textbf{B}).
    With DG discretization the problem unknowns are represented
    in a \emph{single} FE spaces posed over the global mesh. Coupling on
    $\Gamma$ requires no special treatment compared to the remaining facets.
    However, the scheme leads to more degrees of freedom (shown in blue markers, \textbf{C}).}
  \label{fig:domain_dofs}
\end{figure}

\mk{
  To derive the DG discretizations of KNP-EMI sub-problems let us first consider
  their continuous variational formulations. To this end, and to simplify
the exposition, we shall keep the assumption of only a single intracellular domain.
We then let $L^2(\Omega_r)$, $r\in\left\{e, i\right\}$ be the Lebesque space 
of square-integrable functions while $H^1(\Omega_r)$ is the 
standard Sobolev space of functions with derivatives up to order one in $L^2(\Omega_r)$. For $u=(u_i, u_e)\in V$ let us
finally define a membrane jump
\begin{equation}\label{eq:jump_m}
\jumpm{u} = u_i|_{\Gamma} - u_e|_{\Gamma}.
\end{equation}
From the trace theorem in then follows that $\jumpm{u}\in L^2(\Gamma)$ for all $u\in V$, see e.g.~\cite{adams2003sobolev}.
}

\subsubsection{Continuous problems}
\label{sec:weak:form}
Multiplying the EMI sub-problem \eqref{eq:phi:split}--\eqref{eq:phi:split:J} with a suitable test function $w\in V$,
integrating over each sub-domain $\Omega_r$ for $r \in \{i,e\}$, performing integration by parts, summing
over the sub-domains, and inserting interface condition~\eqref{eq:IM:new} yield the following
weak form: Given $c_k^{n-1} \in V$
and $\phi_M^{n-1} \in L^2(\Gamma)$  at time level $n-1$, find the
electrical potential $\phi^n \in V$ at time level $n$ such that:
\begin{align}
\label{eq:emi:weak}
a_{\phi}(\phi^n, w) = l_{\phi}(w) \quad \forall w \in V,
\end{align}
where
\begin{align*}
    a_{\phi}(\phi^n, w) &=\sum_{r \in \{i,e\}} \int_{\Omega_r} \kappa \nabla \phi^n \cdot \nabla w
    + C \int_{\Gamma_m} \jumpm{\phi^n} \jumpm{w}, \\
    l_{\phi}(w) =  
     &- \sum_{r \in \{i,e\}} \int_{\Omega_r} \sum_{k \in {K}} z_k (D_k) \nabla c_k^{n-1} \cdot \Grad w 
    + C \int_{\Gamma_m} f \jumpm{w}
    \\
    &+ F \int_{\partial \Omega}  \sum_{k \in {K}} z_k {\Jvec_k^n} \cdot \nvec w
\end{align*}
with $\kappa = \kappa^{n-1} = F \sum_{k \in {K}} z_k^{2} D_k  \psi
c_k^{n-1}$ and
interface data $C = \frac{C_M}{\Delta t}$ and $f$ given
by~\eqref{eq:f} in the case of a passive membrane model  or~\eqref{eq:f:new} in the case of an
active membrane model (cf.~\Cref{sec:split:active}). \mk{We observe that $a_{\phi}$
  is invariant to the ordering/sign in definition of the jump operator \eqref{eq:jump_m}.
}

Similarly, we obtain the following
weak form of the KNP sub-problem by multiplying~\eqref{eq:con:split}--\eqref{eq:con:split:J} with
a suitable test function, integrating over each sub-domain $\Omega_r$ for $r \in \{i,e\}$, integration
by parts, summing over the sub-domains, and inserting interface conditions~\eqref{eq:mem:conc:i:new}--\eqref{eq:mem:conc:e:new}:
Given $\phi^n \in V$ at time level $n$, for each $k \in \tilde{K}$
find the concentrations $c_k^n \in V$ at time level $n$ such that:
\begin{align}
\label{eq:knp:weak}
\frac{1}{\Delta t}\int_{\Omega} c_k^n v_k  + a_{c_k}(c_k^n, v_k) = \frac{1}{\Delta t} \int_{\Omega}  c_k^{n-1} v_k + l_{c_k}(v_k) \quad \forall v_k \in V,
\end{align}
where
\begin{align*}
\begin{split}
    a_{c_k}(c_k^n, v_k) &= \sum_{r \in \{i,e\}}  \int_{\Omega_r} (D_k \nabla c_k^n \cdot \nabla v_k 
    + z_k \psi D_k c_k^n  \nabla \phi_{r}^n \cdot \nabla v_k), 
    \end{split}\\
    \begin{split}
    l_{c_k}(v_k) = &- \int_{\Gamma_m} ((\jumpm{C_k} \avg{v_k}
    + \avg{C_k} \jumpm{v_k}) \jumpm{\phi} - \jumpm{C_k g_k v_k})
    -  \int_{\partial \Omega} (\Jvec_k \cdot \nvec) v_k
\end{split}
\end{align*}
with interface data $C_k = C_k^{n-1} = \frac{\alpha_{k}C_M}{F z_k \Delta t}$ and $g_k$ given by~\eqref{eq:g}
in the case of a passive membrane model or by~\eqref{eq:g:new} in the case of
active membrane mechanism governed by ODEs (cf.~\Cref{sec:split:active}).

\subsubsection{DG discretization}
\label{sec:weak:form:DG}
\mk{
  Let $\mathcal{T}^h$ be a triangulation of $\Omega$ which conforms to
  the interface $\Gamma$ in the sense that for any element $E\in\mathcal{T}^h$
  the intersection of $\Gamma$ with $\partial E$ is either a vertex or
  an entire face of the element (or an entire edge if $d=3$), cf. \Cref{fig:domain_dofs}.
  Here $h>0$ represents a characteristic mesh size.
  We let $\mathcal{F}^h$ be the collections of all facets of $\mathcal{T}^h$
  and $\mathcal{F}^h_m=\Gamma\cap\mathcal{F}^h$, $\mathcal{F}^h_e=\partial\Omega\cap\mathcal{F}^h$,
  $\mathcal{F}^h_i=\mathcal{F}^h\setminus(\mathcal{F}^h_m\cup \mathcal{F}^h_e)$ shall be respectively, the membrane, exterior and (non-membrane) interior facets.
}

\mk{
  With the conforming mesh
  the continuous weak formulations~\eqref{eq:emi:weak} and~\eqref{eq:knp:weak}
  can be readily discretized by $H^1$-conforming elements (e.g, continuous
  Lagrange elements), posed on meshes $\mathcal{T}^h\cap\Omega_r$, see
  \cite{huynh2023convergence, benedusi2024EMI}. We note that non-conforming
  meshes can be handled by CutFEM techniques \cite{berre2023cut}.
  Here, discretization by DG elements shall be pursued.}

\mk{
  The discrete approximations of \eqref{eq:emi:weak} and~\eqref{eq:knp:weak}
  are obtained by applying standard
  DG techniques to handle the operators in the sub-domains' interiors. In particular,
  the interface terms on $\Gamma$ will remain unchanged from the continuous problem.
  Given the polynomial degree $p \geq 1$, let $\mathbb{P}_p(E)$ be the space of polynomials of degrees
  up to $p$ on $E$.
  We denote by $V^h_p$ the space of functions in $L^2(\mathcal{T}^h)$ whose
  restriction to each element $E$ is in $\mathbb{P}_p(E)$. For $u\in V^h_p$ and any facet $e\in\mathcal{F}^h$
  shared by elements $E_1$ and $E_2$
  we define respectively the jump and average operators
  \begin{equation}\label{eq:jump}
    \jump{u} = u_1 - u_2 , \quad \avg{u} = \frac{u_1 +
      u_2}{2} \quad \text{on } e \in \mathcal{F}_i^h\cup\mathcal{F}^h_m,
  \end{equation}
  where $u_1 =: u|_{E_1}$ and $u_2 =: u|_{E_2}$. Note that the jump
  operator \eqref{eq:jump} extends the membrane jump \eqref{eq:jump_m}
  to all the interior facets.
}
  


\mk{
  Applying the symmetric interior penalty method on the diffusive term with respect to the unknown $\phi$ and with a penalty parameter
  $\beta>0$ (see e.g.~\cite{arnold2002unified} or~\cite{riviere2008discontinuous}),  the discrete weak form of~\eqref{eq:emi:weak} reads:
  Given $c_k^{n-1} \in V^h_p$
  and $\phi_M^{n-1} \in L^2(\Gamma)$ at time level $n-1$, find the
  electrical potential $\phi \in V^h_p$ at time level $n$ such that:
\begin{align}
    \label{eq:emi:weak:dg}
    a_{\phi}^{h}(\phi^n, w) = l^{h}_{\phi}(w) \quad \forall w \in V^h_p,
\end{align}
where}\mk{
\begin{align*}
        a_{\phi}^h(\phi^n, w) &= \sum_{E \in \mathcal{T}^h} \int_E \kappa \nabla \phi^n \cdot 
            \nabla w 
            + C \sum_{e\in\mathcal{F}_i^h}\int_{e}  \jump{\phi^n} \jump{w} \\
            &- \sum_{e\in\mathcal{F}_m^h}\int_{\Gamma_e}
            (\avg{\kappa \nabla \phi^n \cdot \nvec_e} \jump{w}
            + \avg{\kappa \nabla w \cdot \nvec_e}  \jump{\phi^n}
            - \frac{\beta}{h_f} \kappa \jump{\phi^n} \jump{w}),
\end{align*}}
and
\mk{
\begin{align*}
        l_{\phi}^h(w) = 
            &- \sum_{E \in \mathcal{T}^h} F \int_E \sum_{k \in {K}} z^k
            D_k \nabla c_k^{n-1} \cdot \nabla w 
            + F \sum_{e\in\mathcal{F}_i^h}\int_{e} \sum_{k \in {K}} z_k  \avg{D_k \nabla
            c_k^{n-1} \cdot \nvec_e} \jump{w} \\
            &+ C \sum_{e\in\mathcal{F}_m^h}\int_{e}  f \jump{w}
            + F \sum_{e\in\mathcal{F}_e^h}\int_{e} \sum_{k \in {K}} z^k ({\Jvec_k^n}
            \cdot \nvec_e) w 
\end{align*}
}
\mk{
with interface data $C = \frac{C_M}{\Delta t}$ and $f$ given
by~\eqref{eq:f}. We remark that $\beta>0$ needs to be chosen large
enough such that \eqref{eq:emi:weak:dg} is positive definite. Furthermore,
we note that the definition/sign in the jump operator \eqref{eq:jump} does
not alter the bilinear form $a_{\phi}^h$. However, for consistency of $l^h_{\phi}$ (with $l_{\phi}$ in \eqref{eq:emi:weak}) we require
that the operator is consistent with the membrane jump $\jumpm{\cdot}$ \eqref{eq:jump_m}
on $\mathcal{F}^h_m$.}

Similarly, we consider the following discrete approximation of
\eqref{eq:con:split}--\eqref{eq:con:split:J}: Given $\phi^n \in
V^h_p$ at time level $n$, for each $k \in \tilde{K}$ find the concentrations $c_k^n
\in V^h_p$ at time level $n$ such that:
\begin{align}
    \label{eq:knp:weak:dg}
    \sum_{E \in \mathcal{T}_h} \frac{1}{\Delta t} \int_{E} c_k^n  v_k + a_{c_k}^{h}(c_k^n, v_k) = \sum_{E \in \mathcal{T}_h} \frac{1}{\Delta t} \int_{E}  c_k^{n-1} v_k + l_{c_k}^{h}(v_k) \quad \forall v_k \in V^h_p,
\end{align}
where
\begin{align*}
\begin{split}
    a_{c_k}^h(c_k^n, v_k) &= 
    \sum_{E \in \mathcal{T}_h} \int_{E} \bigl( D_k \nabla c_k^n \cdot \nabla v_k
    + z_k \psi D_k c_k^n  \nabla \phi_{r} \cdot \nabla v_k\bigr)\\    
    &-\sum_{e\in\mathcal{F}_i^h}\int_{e} (\avg{D_k \nabla c_k^n \cdot \nvec_e}  \jump{v_k} +  \avg{D_k \nabla v_k \cdot  \nvec_e} \jump{c_k^n}
    - \frac{\gamma}{h_f} D_k \jump{c_k^n} \jump{v_k}) \\
    &-z_k \psi  \sum_{e\in\mathcal{F}_i^h}\int_{e} 
    (\avg{D_k c_k^n \nabla \phi} - \frac{|D_k \nabla \phi \cdot
    \nvec_e|}{2}\jump{c_k^n}) \jump{v_k}, 
    \end{split}\\
    \begin{split}
    l_{c_k}^h(v_k) = 
    &- \sum_{e\in\mathcal{F}_m^h}\int_{e} ((\jump{C_k} \avg{v_k}
    + \avg{C_k} \jump{v_k}) \jump{\phi} - \jump{C_k g_k v_k}) -  \sum_{e\in\mathcal{F}_e^h}\int_{e} (\Jvec_k \cdot \nvec) v_k
\end{split}
\end{align*}
with interface data $C_k = C_k^{n-1} = \frac{\alpha_{k}C_M}{F z_k \Delta t}$ and $g_k$
given by~\eqref{eq:g}. Here, we have
applied a symmetric interior penalty method on the diffusive term with penalty
parameter $\gamma$, and the advective term is up-winded (see e.g.~\cite{arnold2002unified} or~\cite{riviere2008discontinuous}). The weak form~\eqref{eq:knp:weak} is obtained by
multiplying~\eqref{eq:con:split} with a suitable test function $v_k \in
V_p^{h}$, integrating over one element $E \in \mathcal{T}_h$,
integration by parts on both the diffusive term and the advective term, summing
over all elements $E \in \mathcal{T}_h$, and inserting interface
conditions~\eqref{eq:mem:conc:i:new}--\eqref{eq:mem:conc:e:new} in the integral
over $\Gamma_m$. 


\subsection{Extension of scheme for active (ODE governed) membrane models}
\label{sec:split:active}
In the active case, where the total ionic current $\Ich = \Ich(\phi_M, \mathbf{s})$ depends on the membrane
potential $\phi_M$ and gating variables $\mathbf{s}$ governed by ODEs of the
form~\eqref{eq:ODE:pot}--\eqref{eq:ODE:gate}, we add an additional first order
Godunov splitting step (see e.g.~\cite{sundnes2007computing}) to the splitting scheme presented in \Cref{sec:time:discretization}. In the first step, we update the membrane potential $\phi_M^n$ at time
step $t^n$ by solving the ODE system~\eqref{eq:ODE:gate}--\eqref{eq:ODE:pot}
with $I_M$ set to zero, using a suitable ODE solver with time step $\Delta
t^{\ast}$. In the second step, we solve for $\phi^n$ and $c_k^n$
in~\eqref{eq:phi:split}--\eqref{eq:con:split:J} (i.e.~solve step I and II) with
$\Ich$ set to zero in interface condition~\eqref{eq:IM:new}. This results in the following new expression 
\begin{align}
\label{eq:f:new}
    f = f^{n-1} = \phi_M^{n-1}
\end{align}
replacing~\eqref{eq:f}.
Setting $\Ich$ to
zero in~\eqref{eq:IM:new} will further affect the interface
conditions~\eqref{eq:mem:conc:i:new}, and~\eqref{eq:mem:conc:e:new}.
Specifically, we obtain the following new expression 
\begin{equation}
   \label{eq:g:new}    
    g_k = g_k^{n-1} =  
    \phi_M^{n-1} - \frac{\Delta t}{C_M \alpha_{k}} \Ichk + \frac{\Delta t}{C_M} \Ich
\end{equation}
replacing~\eqref{eq:g}. An outline of the complete solution algorithm, for the case with an active membrane model, can be found in Algorithm~\ref{alg:cap:active}.

\begin{algorithm}
\caption{Solution algorithm for the KNP-EMI system with active membrane}
    \label{alg:cap:active}
\begin{algorithmic}
    \State set $c_k^0$ for $k \in \tilde{K}$ and  $\phi_M^0$ 
\For{$n = 1, \dots, N$}
    \State set $\phi_M^{m-1} = \phi_M^{n-1}$ and solve ODEs:
    \For{$m = 1, \dots, M$}
        \State solve ODEs
        \State $\phi_M^{m-1} \gets \phi_M^m$
    \EndFor
    \State set $\phi_M^{n-1} = \phi_M^m$ and solve PDEs:
        \State find $\phi^n$ by solving~\eqref{eq:emi:weak:dg} with interface data~\eqref{eq:f:new}
        \State find $c_k^n$ for $k \in \tilde{K}$ by solving~\eqref{eq:knp:weak:dg} with interface data~\eqref{eq:g:new}
        \State $\phi_M^{n-1} \gets (\phi_i^n - \phi_e^n)$
        \State $c_k^{n-1} \gets c_k^{n}$ for $k \in \tilde{K}$ (and thus updating $\alpha_k, E_k, \Ichk$ and $\kappa$)
        \EndFor
\end{algorithmic}
\end{algorithm}

\section{Solvers}
\label{sec:solver}
The numerical scheme presented in \Cref{sec:numerical:scheme} results in an algorithm where three discrete sub-problems must be solved:~(i) an EMI problem governing the electrical potentials, (ii) a series of advection diffusion (KNP) problems governing ion transport, and (iii) a system of ODEs defined on the mesh facts representing the lower dimensional (membrane) interfaces between the sub-regions. 

\subsection*{EMI sub-problem}
\mk{
The DG discretization \eqref{eq:emi:weak:dg} of EMI problem leads to a linear
system $A_{\phi} x_{\phi} = b_{\phi}$ for the expansion coefficients $x_{\phi}\in\mathbb{R}^m$
of the unknown potential $\phi^n$ with respect to the basis of $V^h_p$, $\text{dim} V^h_p=m$.
As noted earlier, for sufficiently large stabilization parameter $\beta$
the problem matrix $A_{\phi}$ is \emph{positive semi-definite} with
vector $z_\phi=1$ in the kernel, i.e. $A_{\phi}z_{\phi}=0$. Solvability of the linear system
then requires that $b_{\phi}$ is orthogonal (in the $l^2$ inner product) to the kernel.
For such compatible $b_{\phi}$, unique solution can be obtained
in the orthogonal complement of $R^m$ with respect to $z_{\phi}$.
}

\mk{
  Due to symmetry and positivity of $A_{\phi}$, and following orthogonalization of $b_{\phi}$,
  the unique $x_{\phi}$ can be obtained by conjugate gradient (CG) method informed about
  the nullspace, see e.g.~\cite{KAASSCHIETER1988265}.  %
  To accelerate convergence of the Krylov solver we will
  consider a preconditioner realized as a single V-cycle of algebraic multigrid
  (AMG)\cite{hypre} 
  applied to the \emph{positive definite} matrix
  $P_{\phi} = A_{\phi} + \alpha M$ where $M$ is the mass matrix of $V^h_p$ and
  $\alpha>0$ is a scaling parameter dependent on the domain.
  We remark that, theoretical analysis of AMG for the EMI sub-problem is an
  active area of research. In particular, using conforming FEM discretizations
  \cite{budisa2023algebraic} 
  prove uniform convergence of aggregation-based AMG with custom smoothers
  to handle the interface. 
  However, with DG discretization,
  assuming that $\Delta t$ scales at most linearly in $h$,
  the interface $\Gamma$
  (or $\mathcal{F}^h_m$) does not present additional challenge compared to
  the remaining facet couplings on $\mathcal{F}^h_i$. Therefore, existing
  AMG algorithms for elliptic problems, e.g.~\cite{antonietti2020algebraic} 
  and references therein,
  could be applied.
}

\subsection*{KNP sub-problem}
\mk{
The discrete KNP sub-problem~\eqref{eq:knp:weak:dg} yields a linear system
$A_{c} x_{c} = b_{c}$. Here, the lower-order term stemming from temporal
discretization ensures\footnote{Even if homogeneous Neumann boundary conditions are applied.}
that the matrix $A_{c}$ is positive-definite. Since the problem is not
symmetric we apply generalized minimal residual method (GMRes)
using single AMG\cite{hypre} V-cycle applied to $A_c$ as the preconditioner.}

\subsection*{ODE solver}
For the ODE step in our splitting algorithm we apply the LSODA method
\cite{hindmarsh2005lsoda} from ODEPACK \cite{odepack}. 

\section{Simulation scenarios and parameters}
In this section, we define five simulations scenarios, with increasing complexity and physiological relevance, that will be used to numerically investigate the discretization scheme and solvers presented in respectively \Cref{sec:numerical:scheme} and~\ref{sec:solver}. We always consider three ion species, namely potassium ($\rm K^+$), sodium ($\rm Na^+$) and chloride ($\rm Cl^-$), and model parameters and initial data given in \Cref{tab:physconst} unless otherwise stated in the text.
 \begin{table}
 \begin{center}
    \begin{tabular}{llll}
    \toprule
        Parameter & Value &  Parameter & Value \\
    \midrule
        $R$ &  8.314 J/(K mol) & $c_{\text{Na},i}^0$ & 12  mM \\
        $T$ &  300  K  & $c_{\text{Na},e}^0$ & 100  mM  \\
        $F$ &  $9.648\cdot 10^4$  C/mol  & $c_{\text{K},i}^0$ & 125 mM \\    
        $C_M$ & $0.01$  F/m
       &  $c_{\text{K},e}^0$ & 4  mM \\
        $D^\text{Na}_r$ & $1.33\cdot 10^{-9}$ m\textsuperscript{2}/s 
        &  $c_{\text{Cl},i}^0$ & 137  mM \\ 
        $D^\text{K}_r$  & $1.96\cdot 10^{-9}$ m\textsuperscript{2}/s 
        &  $c_{\text{Cl},e}^0$ & 104  mM \\
        $D^\text{Cl}_r$ &  $2.03\times 10^{-9}$  m\textsuperscript{2}/s &
        $\phi_M^0$ & $-67.74 \times 10^{-3}$ V \\
        $g_{\text{leak,Na}}$ & 1.0  S/m\textsuperscript{2} & 
        $\Delta t$ & $1.0\times 10^{-4}$ s \\
        $g_\text{leak,K}$  & 4.0 S/m\textsuperscript{2}  
        & $\Delta t^{\ast}$ & $1.0\times 10^{-4}$ s  \\
        $g_{\rm syn}$ &  40 S/m$^2$ & 
        $\beta, \gamma$ & $20\times d \times p$
        \\
        $\tau$ &  $0.02$ s & & \\ 
   \bottomrule
    \end{tabular}
    \caption{\label{tab:physconst} Physical and numerical parameters and initial values
    used in the simulations. All units are reported
    in SI base units. The Hodgkin-Huxley parameters are taken from~\cite{ellingsrud2020finite}. The penalty parameter $\beta$ arising from the DG discretization depends on the geometrical dimension (d) and the polynomial degree of the element (p). The ODE solver in our implementation is adaptive with respect to the timestep, and $\Delta t^{\ast}$ is the maximum ODE timestep.} \end{center}
\end{table}
 
\subsection{Model A: Smooth manufactured solutions}
To evaluate the numerical accuracy of the discretization scheme presented in
\Cref{sec:numerical:scheme}, we construct two analytical solutions using the method of
manufactured solutions: one that is constant in time (\eqref{eq:MMS}, to assess
spatial accuracy), and one that is linear in space (\eqref{eq:MMS:time}, to
assess temporal accuracy). In both cases, we consider a two-dimensional domain
$\Omega = \Omega_i \cup\Omega_e = [0, 1] \times [0, 1]$, with one intracellular sub-domain
$\Omega_i = [0.25, 0.75 ] \times [0.25, 0.75]$. To assess the numerical errors from the spatial discretization scheme, we let the analytical solutions to \eqref{eq:concentration:org}--\eqref{eq:phi:org} be given by: 
\begin{equation}
\begin{aligned}
    \label{eq:MMS}
    c_{\textrm{Na}, i} &= 0.7 + 0.3\sin(2\pi x)\sin(2\pi y)
    && \qquad \text{in } \Omega_{i}, \\
    c_{\textrm{Cl}, i} &= 0.3 + 0.4\cos(2\pi x)\sin(2\pi y)
    && \qquad \text{in } \Omega_{i}, \\
    \phi_i &= \cos(2\pi x) \cos(2\pi y)
    && \qquad \text{in } \Omega_{i}, \\
    c_{\textrm{Na}, e} &= 0.7 + 0.2\cos(2\pi x)\cos(2\pi y)
    && \qquad \text{in } \Omega_{e}, \\
    c_{\textrm{Cl}, e} &= 0.3 + 0.8\sin(2\pi x)\cos(2\pi y)
    && \qquad \text{in } \Omega_{e}, \\
    \phi_e &= \sin(2\pi x)\sin(2\pi y)
    && \qquad \text{in } \Omega_{e}.
\end{aligned}
\end{equation}
\mk{We then obtain
a series of uniformly refined meshes of the domain by first subdividing
$\Omega$ into $n\times m$ rectangles each of which is then split into two triangles.
We let $n = m \in \left\{16, 32, 64, 128, 256\right\}$.} We further let $\Delta t = 1 \times 10^{-10}$ and evaluate the errors at $t = 2 \times 10^{-10}$. To assess the numerical errors from the temporal discretization and PDE splitting scheme, we let the analytical solutions to \eqref{eq:concentration:org}--\eqref{eq:phi:org} be given by:
\begin{equation}
\begin{aligned}
    \label{eq:MMS:time}
    c_{\textrm{Na}, i} &= 1 + x + y + 0.3\cos(2 \pi t)
    && \qquad \text{in } \Omega_{i}, \\
    c_{\textrm{Cl}, i} &= 1 + x + y + 0.2\cos(2 \pi t)
    && \qquad \text{in } \Omega_{i}, \\
    \phi_i &= 1 + x + y
    && \qquad \text{in } \Omega_{i}, \\
    c_{\textrm{Na}, e} &= 1 + x + y + 0.5\sin(2 \pi t)
    && \qquad \text{in } \Omega_{e}, \\
    c_{\textrm{Cl}, e} &= 1 + x + y + 0.6\sin(2 \pi t)
    && \qquad \text{in } \Omega_{e}, \\
    \phi_e &= 1 + x + y
    && \qquad \text{in } \Omega_{e}.
\end{aligned}
\end{equation}
We mesh the domain with $n = m = 64$ and consider first order DG elements. We initially let
$\Delta t = 1 \times 10^{-6}$ and half the time step on refinement. The errors are
evaluated at $t = 2 \times 10^{-6}$.

\subsection{Model B: Idealized 2D axon with active membrane model} 
\label{sec:model:B}
We next consider a model scenario with active membrane mechanisms depending on gating variables governed by ODEs. The domain is defined as $\Omega = \Omega_i \cup
\Omega_e = [0, 62] \times [0, 4]$ $\mu$m, with one ICS domain $\Omega_i =
[1, 61] \times [1, 3]$ $\mu$m, representing an idealized axon embedded in ECS. The membrane model, i.e.~the ion channel currents $\Ichk$, are governed by the Hodgkin-Huxley model adapted to take into account explicit representation of varying ion concentrations (see e.g.~\cite{ellingsrud2020finite}). 
An action potential is induced every 20 ms throughout the simulations by applying the following
synaptic input current model:
\begin{equation}
\label{eq:Isyn}
I_{\rm syn}(\vec{x},t) = g_{\rm syn} f(\vec{x}) e^{-\frac{t}{\tau}}(\phi_M - E_{\rm Na}),    
\end{equation}
where $\tau$, $g_{\rm syn}$ and $E_{\rm Na}$ respectively denote the synaptic time constant and strength, and the $\rm Na^{+}$ Nernst potential. The function $f$ will vary with the geometry of interest, and for model B we define:
\begin{equation}
\label{eq:f:source}
    f(\vec{x}) =
    \begin{cases}
        1 & \text{if } x \leq  1 \mu\rm m\\
        0 & \text{else} 
    \end{cases}
    \qquad \vec{x}=(x, y)\in \Gamma.
\end{equation}
We consider four different meshes of the 2D geometry with increasing size that have respectively
3968, 15872, 63488, 253952 mesh cells. In the numerical experiments using model B, the EMI and
KNP sub-problems are discretized with first order DG elements.

\subsection{Model C: Idealized 3D axons with active membrane model}
\label{sec:model:C}
For the third model scenario, we consider the following 3D geometry representing four idealized axons embedded in ECS:
$\Omega = \Omega_{i^{1}} \cup \cdots \cup \Omega_{i^{4}} \cup \Omega_e = [0, 32] \times [0, 0.9] \times [0, 0.9]$ $\mu$m, where 4 cuboidal
cells of size $6 \times 0.2 \times 0.2$ $\mu$m are placed
uniformly throughout $\Omega$ and where the distance between the cells is $0.1$ $\mu$m. The synaptic input current is given by~\eqref{eq:Isyn} with:
\begin{equation}
\label{eq:f:C}
    f(\vec{x}) =
    \begin{cases}
        1 & \text{if } x \leq  20 \mu\rm m\\
        0 & \text{else} 
    \end{cases}
    \qquad \vec{x}=(x, y, z)\in\Gamma.
\end{equation}
We consider two meshes of the 3D geometry with respectively 3968 and 15872 number of mesh cells. In the numerical experiments using model C, the EMI and KNP sub-problems are discretized with first order DG elements.
The membrane model is the same as in Model B. 

\subsection{Model D: Morphologically realistic neuron with spatially varying membrane model}
\label{sec:model:D}
Next, we consider a model with a physiologically realistic cellular geometry representing a
pyramidal neuron in the cortex of a dog, based on a digitally reconstructed neuron from the
neuromorpho database\footnote{\url{https://neuromorpho.org/}}, 
embedded in a bounding box representing the ECS. \mk{Using the centerline/skeleton
representation of the neuron geometry, we obtain the computational
mesh in two steps. First, AnaMorph \cite{anamorph} is used to reconstruct the neuron surface
as a surface embedded in 3D. The resulting surface representation is then meshed by fTetWild \cite{hu2020fast}.
}
The neuron in stimulated at the
tip of the dendrites, i.e.~the synaptic input current is given by~\eqref{eq:Isyn} with:
\begin{equation}
\label{eq:f:D}
    f(\vec{x}) =
    \begin{cases}
        1 & \text{if }
        x \leq -125 \mu{\rm m}, y \geq 140\mu{\rm m},  
        z \leq -80 \mu{\rm m}
        \\
        0 & \text{else}
    \end{cases}
    \quad \vec{x}=(x, y, z)\in\Gamma.
\end{equation}
We apply the same Hodgkin-Huxley model as in model B (cf.~\Cref{sec:model:B}) at the soma and axon membranes, and a passive model of the form~\eqref{eq:Ichk} with $g_{\rm Na} = g_{\text{leak,Na}}$ and $g_{\rm K} = g_{\text{leak,K}}$ given by \Cref{tab:physconst}, and $g_{\rm Cl} = 0$.

\subsection{Model E: 
Dense reconstruction of the visual cortex}
\label{sec:model:E}
In the final model scenario, we consider three meshes of a dense reconstruction of the mouse visual cortex with ECS and various cellular structures\footnote{\url{https://zenodo.org/records/8356333}, \url{https://www.microns-explorer.org/cortical-mm3}}. The three meshes represent cubes of reconstructed tissue with dimension $5 \times 5 \times 5 \mu \rm m$ containing respectively 5, 50 and 100 brain cells.

\subsection{Peclet number}
\label{sec:methods:peclet}
In standard advection diffusion problems, the ratio between advective and diffusive contributions to transport is typically quantified by the Peclet number:
\begin{align}
\label{eq:peclet}
    \textrm{Pe} = \frac{L u}{D},
\end{align}
where $L$ is the length scale, $u$ denotes the velocity, and $D$ denotes the
diffusion coefficient. In the KNP-EMI system, the advection in the system is not driven by a
standard velocity field, but rather by drift,
i.e.~$z_k \frac{FD}{RT} \nabla \phi$ (cf.~\eqref{eq:fluxJ:org}). Replacing the velocity field
$u$ by the drift term in~\eqref{eq:peclet} we get
\begin{align}
    \textrm{Pe} = \frac{L z_k \frac{FD}{RT} \nabla \phi}{D} = \frac{L z_k F \nabla \phi}{RT},
\end{align}
i.e.~the diffusion coefficients $D$ cancel out.
As such, we introduce the following measure for quantifying the relative contribution of advection and diffusion in the KNP-EMI system:
\begin{align}
    \label{eq:peclet:new}
    \frac{|D \nabla c_k| - |z_k c \frac{FD}{RT} \nabla \phi|}{\rm{max} (|D \nabla c_k|, \epsilon)}.
\end{align}
This function is positive where diffusion dominates, and negative where drift
dominates. The denominator ensures that the expression is well defined in the
case of a zero concentration gradients.

\section{Results}
We here present results from numerical experiments using the KNP-EMI framework and the solution strategy outlined above. To study accuracy and convergence of the scheme we apply model A. We investigate robustness and scalability of the solver by using models B and C, before assessing how the solver behaves in the case of realistic brain tissue geometries with model D and E. Finally, we assess the contribution from advective and diffusive forces for physiologically relevant scenarios. All the numerical experiments, except for the parallel scaling experiments, are run in serial on a laptop with 8 11th Gen Intel 2.8 GHz cores with 32 GiB memory. The parallel scaling experiments are run on the SAGA supercomputer where each computing node is equipped with 40 Intel Xeon-Gold 6138 2.0 GHz cores with 192 GiB memory
each\footnote{\url{https://documentation.sigma2.no/hpc_machines/saga.html}}.

\subsection{Implementation and solver settings}
The KNP-EMI solver outlined in \Cref{sec:numerical:scheme} and \Cref{sec:solver} is implemented using FEniCS~\cite{alnaes2015fenics, logg2012automated}, an open source computing platform for solving PDEs with the finite element method, with PETSc~\cite{petsc-user-ref} as the linear algebra backend. The discrete EMI and KNP sub-problems are solved using preconditioned CG and GMRes solvers 
respectively. Implementations of both CG and GMRes are provided by PETSc \cite{petsc-user-ref}.
 For each timestep $n$, we use the solution at $n-1$ as the initial guess for both the CG and GMRes solvers to reduce the number of iterations before convergence. To determine convergence of the solution, we use a relative tolerance of $10^{-5}$ and $10^{-7}$ for respectively CG and GMRes in all the numerical experiments presented below. The bounds are chosen such that we do not observe any numerical artifacts in the solutions for models B--D.
 We use MUMPs \cite{MUMPS:1} as the direct solver. For the further details about the solver setup we refer to the associated software repository~\cite{ellingsrud_2024_10953504}. 


\subsection{Temporal and spatial errors and convergence}
Using Model A, we analyze the convergence rates for the approximations of all
solution variables under refinement in space and time. Based on properties of
the approximation spaces, the theoretically optimal spatial rate of convergence
in the $L^2$-norm is $p+1$ where $p$ is the polynomial degree. Our numerical
findings are in agreement with the expected optimal rates: we observe
second order convergence in the $L^2$-norm for the approximation of
concentrations and the electrical potential when discretizing by linear elements
(\Cref{tab:con:rates:space:1}), and third order convergence when quadratic DG
elements are used (\Cref{tab:con:rates:space:2}).

The optimal theoretical rate of the temporal PDE discretization scheme,
consisting of a first order time stepping scheme and a first order PDE operator splitting scheme,
is $1$. We observe first order convergence in the $L^2$-norm for the approximations of the concentrations and the electrical potential
(\Cref{tab:con:rates:time}). 

\begin{table}
 \begin{center}
    \begin{tabular}{lllll}
    \toprule
        $h$ & $p$ & $||e_{[a]}||_{L^2}$ (r) & $||e_{[b]}||_{L^2}$ 
        (r) & $||e_{\phi}||_{L^2}$ (r)  \\
    \midrule
        0.353 & 1 & 4.78e-2 --     & 4.78e-2 --     & 1.05e-2 -- \\
        0.177 & 1 & 1.38e-2 (1.79) & 1.38e-2 (1.80) & 3.36e-2 (1.64) \\
        0.088 & 1 & 3.56e-3 (1.95) & 3.56e-3 (1.95) & 9.19e-3 (1.87) \\
        0.044 & 1 & 8.98e-4 (1.99) & 8.99e-4 (1.99) & 2.36e-3 (1.96) \\
        0.022 & 1 & 2.25e-4 (2.00) & 2.25e-4 (2.00) & 5.93e-4 (1.99) \\
        0.011 & 1 & 5.61e-5 (2.00) & 5.68e-5 (2.00) & 1.48e-4 (2.00) \\
   \bottomrule
    \end{tabular}
    \caption{\label{tab:con:rates:space:1} 
    $L^{2}$-error norms and associated convergence rates for concentrations and potentials
    during refinement in space with fixed time step ($\Delta t = 1.0 \times 10^{-10}$) using elements with polynomial degree 1 and a stationary spatially smooth manufactured solution.
    } \end{center}
\end{table}
\begin{table}
 \begin{center}
    \begin{tabular}{lllll}
    \toprule
        $h$ & $p$ & $||e_{[a]}||_{L^2}$ (r) & $||e_{[b]}||_{L^2)}$ 
        (r) & $||e_{\phi}||_{L^2}$ (r)  \\
    \midrule
        0.353 & 2 & 6.48e-3 --   & 6.48e-3 --       & 8.45e-3 -- \\
        0.177 & 2 & 8.58e-4 (2.92) & 8.58e-4 (2.92) & 8.77e-3 (3.27) \\
        0.088 & 2 & 1.08e-4 (2.98) & 1.08e-4 (2.98) & 1.02e-4 (3.10) \\
        0.044 & 2 & 1.36e-5 (2.99) & 1.36e-5 (2.99) & 1.25e-5 (3.03) \\
        0.022 & 2 & 1.71e-6 (3.00) & 1.71e-6 (3.00) & 1.55e-6 (3.01) \\
        0.011 & 2 & 2.13e-7 (3.00) & 2.13e-7 (3.00) & 1.94e-7 (3.00) \\
   \bottomrule
    \end{tabular}
    \caption{\label{tab:con:rates:space:2}
    $L^{2}$-error norms and associated convergence rates for concentrations and potentials
    during refinement in space with fixed time step ($\Delta t = 1.0 \times 10^{-10}$) using elements with polynomial degree 2 and a stationary spatially smooth manufactured solution.
    } \end{center}
\end{table}

\begin{table}
 \begin{center}
    \begin{tabular}{llll}
    \toprule
        $\Delta t$ & $||e_{[a]}||_{L^2(\Omega)}$ (r) & $||e_{[b]}||_{L^2(\Omega)}$ 
        (r) & $||e_{\phi}||_{L^2(\Omega)}$ (r)  \\
    \midrule
        5.00e-3 & 3.50e-3   -- & 2.40e-3 --   & 8.95e-4 -- \\
        2.50e-3 & 1.99e-3 (0.82) & 1.44e-3 (0.74) & 5.95e-4 (0.59) \\
        1.25e-3 & 1.06e-3 (0.90) & 7.96e-4 (0.85) & 3.39e-4 (0.81) \\
        6.25e-4 & 5.49e-4 (0.95) & 4.18e-4 (0.93) & 1.80e-4 (0.92) \\
        3.13e-4 & 2.79e-4 (0.98) & 2.14e-4 (0.97) & 9.21e-5 (0.96) \\
        1.56e-4 & 1.40e-4 (1.00) & 1.08e-4 (0.99) & 4.65e-5 (0.99) \\
        7.81e-5 & 7.03e-5 (1.00) & 5.40e-5 (1.00) & 2.33e-5 (1.00) \\
   \bottomrule
    \end{tabular}
    \caption{\label{tab:con:rates:time} 
    $L^{2}$-error norms and associated convergence rates for concentrations and potentials during
    refinement in time with fixed mesh size ($n = m = 16$) obtained by using a temporally smooth manufactured solution that is linear in space and elements with polynomial degree 1.
    } \end{center}
\end{table}


\subsection{Scalability and robustness of solver}
\begin{figure}
    \centering
    \includegraphics[width=1.\linewidth]{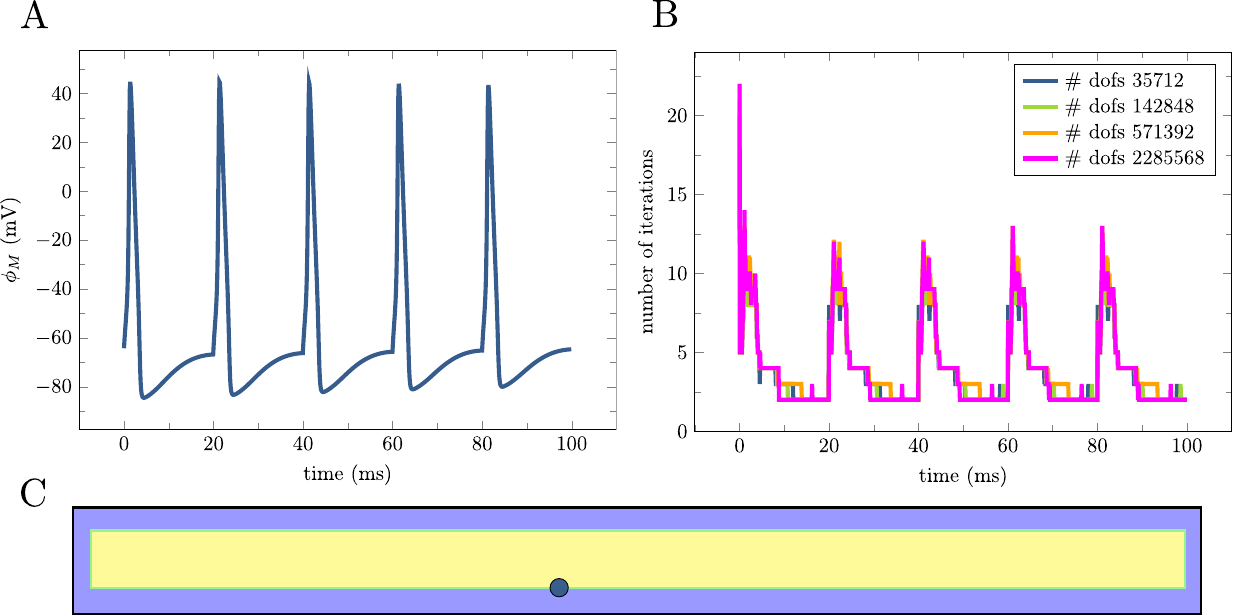}
    \label{fig:itcount:2D}
\vspace{-20pt}    
    \caption{Robustness of the preconditioner with respect to mesh size
    for model B (idealized 2D axon). The figure displays the
    temporal evolution of the membrane potential at point
    $(25, 1)$ $\mu$m (\textbf{A}), the total number of iterations over time during refinement in
    space for the full KNP-EMI system (\textbf{B}), and an illustration
    of the 2D domain with ICS in yellow and ECS in purple (\textbf{C}).}
\end{figure}

\begin{figure}
    \centering
    \includegraphics[width=1.\linewidth]{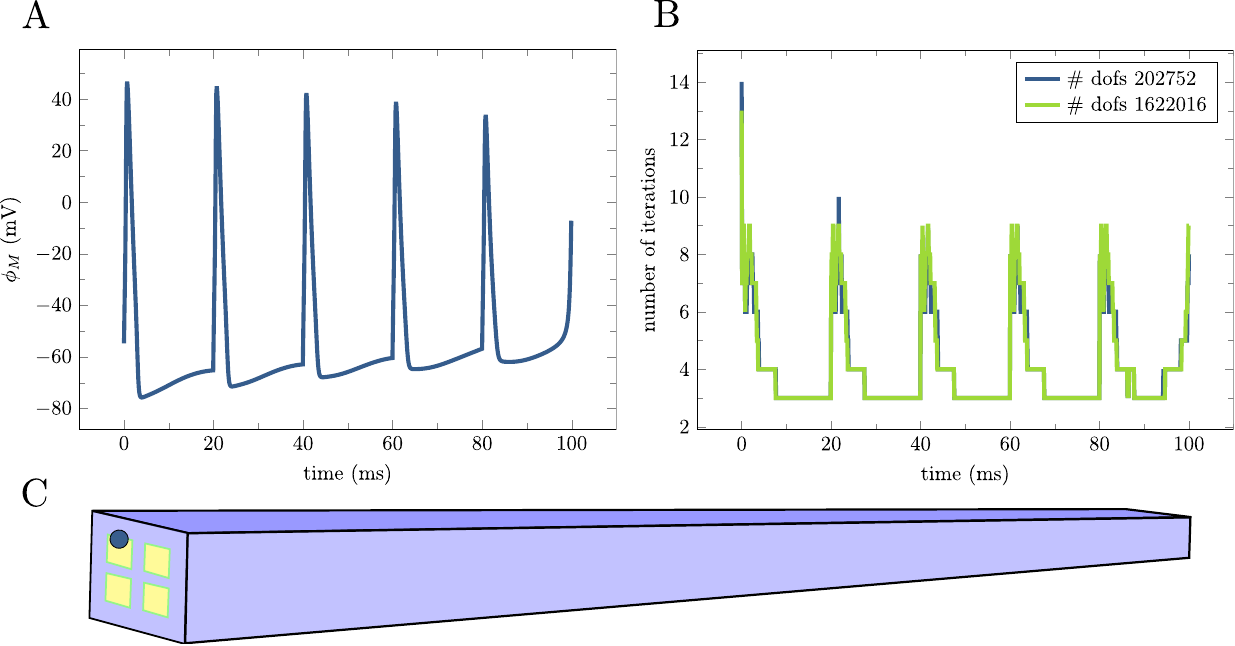}
    \label{fig:itcount:3D}
    \vspace{-20pt}
    \caption{Robustness of the preconditioner with respect to mesh size for model C
    (idealized 3D axons). The figure displays the temporal
    evolution of the membrane potential at point $(25, 0.3, 0.4)$ $\mu$m (\textbf{A}),
    the total number of iterations over time during refinement in space for the full KNP-EMI
    system (\textbf{B}), and an illustration of the 3D domain with ICS in
    yellow and ECS in purple (\textbf{C}).}
\end{figure}

Our motivation for solving the linear systems arising from the discretization of
the KNP-EMI system using a preconditioned iterative solver is to enable
large-scale 3D simulations. To assess the robustness of the solver with respect
to the mesh resolution, we run model B (2D idealized geometry,
\Cref{sec:model:B}) and model C (3D idealized geometry,
\Cref{sec:model:C}) during refinement in space. We observe that the
number of iterations before convergence is stable during refinement in space
for both model B (\Cref{fig:itcount:2D}) and model C
(\Cref{fig:itcount:3D}). As the initial guess in the iterative solver is taken
to be the solution from the previous global time step, we observe a peak in the
number of iterations in the first time step where the initial guess for the
electrical potential is zero. The number of iterations vary with the dynamics
of the system: the number of iterations reaches maximum at respectively $14$ and $9$ when
the action potentials peaks, (at t=20, 40, 60, 80 s), and drops to respectively
$2$ and $3$ when the neuron is silent for model B and model C. On average, the
number of iterations are $3.8$ for mode B and $4.1$ for model C with the
highest mesh resolutions. 
 Note that the slight, but persistent,
depolarization of the membrane potential in model C is due to shifts in
concentration shifts (\Cref{fig:itcount:3D}A).

We next assess the memory usage and CPU timings of the preconditioned
iterative solver for model C, and further compare it to that of a direct LU
solver. As expected, the preconditioned solver has a lower maximum memory
usage than the direct solver ($\sim 50 \%$, \Cref{tab:direct:iterative}).
The assembly times for the solvers are comparable, whereas the time to solve
the linear system is $1.29$ seconds on average for the iterative solver and
$3.69$ seconds on average for the direct solver. 

%

\begin{table}
    \begin{center}
    \begin{tabular}{l|l|lcccc}
    \toprule
        Problem & Size &Solver & Memory &T$_{\rm PDE}^{\rm A}$ & T$_{\rm PDE}^{\rm S}$ & T$_{\rm PDE}$ \\
    \midrule
        \multirow{2}{*}{knp-emi} & \multirow{2}{*}{248832} & LU & 1260 & 0.52 &
        3.69 & 4.21 \\
                                         \cline{3-7} && CG/GMRes (AMG) & 614 &
                                         0.58 & 1.29 & 1.67  \\
    \midrule
        \multirow{2}{*}{emi} & \multirow{2}{*}{62208} & LU & 1260 & 0.19 & 1.03 & 1.22 \\
                                     \cline{3-7} && CG (AMG) & 614 & 0.25 &
                                     0.21 & 0.26 \\
    \midrule
        \multirow{2}{*}{knp} & \multirow{2}{*}{186624} &
        LU & 1260 & 0.33 & 2.66 & 2.99
                                  \\ \cline{3-7} && GMRes (AMG) & 614 & 0.33 &
                                  1.08 & 1.41 \\
    \bottomrule
\end{tabular}
    \caption{\label{tab:direct:iterative} Comparison of average CPU timings and
        memory usage for the preconditioned iterative solver (CG/GMRes + AMG)
        and a direct solver (LU) for model C (idealized 3D axons). Size:~linear
        system size (number of degrees of freedom), Memory: maximal memory
        (MiB) usage of simulation relative to baseline. ${\rm T}_{\rm PDE}^{\rm
        A}$:~ Average CPU time (s) for finite element assembly for one global
        time step, ${\rm T}_{\rm PDE}^{\rm S}$:~Average CPU time (s) for linear
        system solution for one global time step, ${\rm T}_{\rm PDE}$:~Average
        total CPU time (s) for PDE simulation for one global time step.}
    \end{center}
\end{table}

\subsection{Parallel scalability}
The splitting schemes presented in \Cref{sec:solver} result in a
solution algorithm where three sub-systems must be solved, namely the EMI and
the KNP sub-problems and the ODE system. We next assess how the CPU times for
solving the three sub-problems scale with the number of processors. 
Specifically, we perform two scaling studies: (i) a strong scaling study were
we consider an increasing number of cores using the setup in model C, and (ii)
a weak scaling study where we increase the mesh resolution and the number of
cores simultaneously, such that the number of degrees of freedom per core stays
constant, using the setup in model B. 

\begin{figure}
    \centering
    \includegraphics[width=\textwidth]{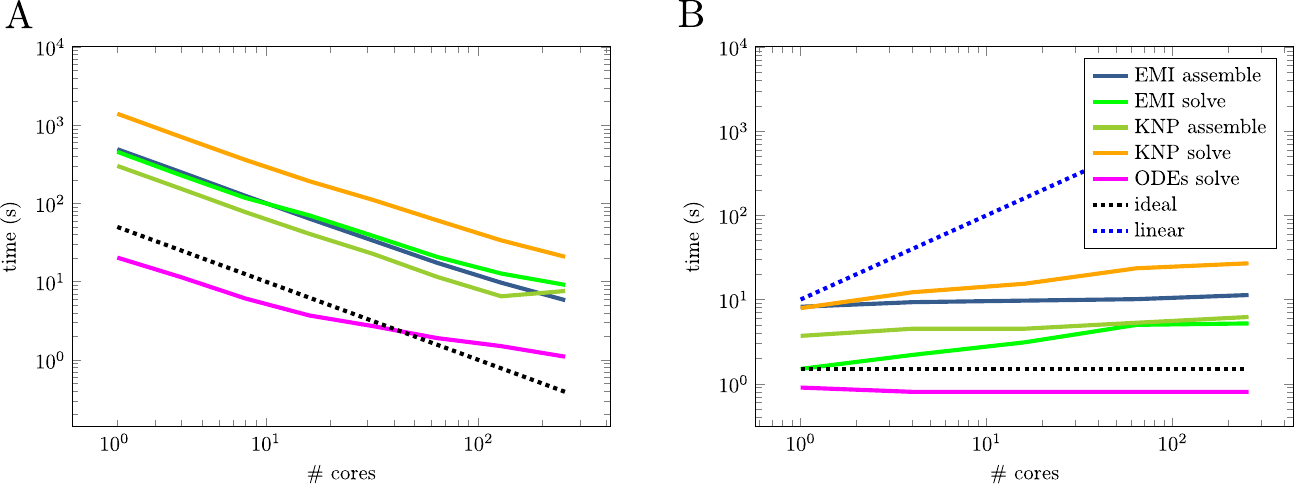}
    \vspace{-20pt}
    \caption{Parallel scalability of assembly and solve times for the EMI and the KNP sub-problems, and for solving the ODEs. Strong scaling data for model C (idealized 3D geometry), with ideal scaling reported (\textbf{A}, log-log plot). Weak scaling data for model B (idealized 2D geometry), with both ideal and linear scaling reported (\textbf{B}, log-log plot). The timings represent total runtime and assembly over 10 timesteps.}
    \label{fig:scaling}
\end{figure}

The CPU time required to assemble and solve both the EMI and the KNP
sub-problems decreases linearly with the number of cores, which is close to
the expected (and ideal) linear scaling (\Cref{fig:scaling}A).
Similarly, the CPU time for the ODEs scales linearly until we reach $\sim 50$
cores, where the CPU time flattens out. 

In the weak scaling study, we observe a sub-linear scaling of the CPU time per
core for both assembling and solving the EMI and the KNP sub-problems, whereas
the CPU time per core for solving the ODE stays constant
(\Cref{fig:scaling}B). As the number of degrees of freedom, and
consequently the size of the matrix, per core is constant, the ideal scaling is constant.  

\subsection{Ion transport is locally dominated by drift}
\label{sec:results:peclet}
\begin{figure}
    \centering
    \includegraphics[width=1.0\linewidth]{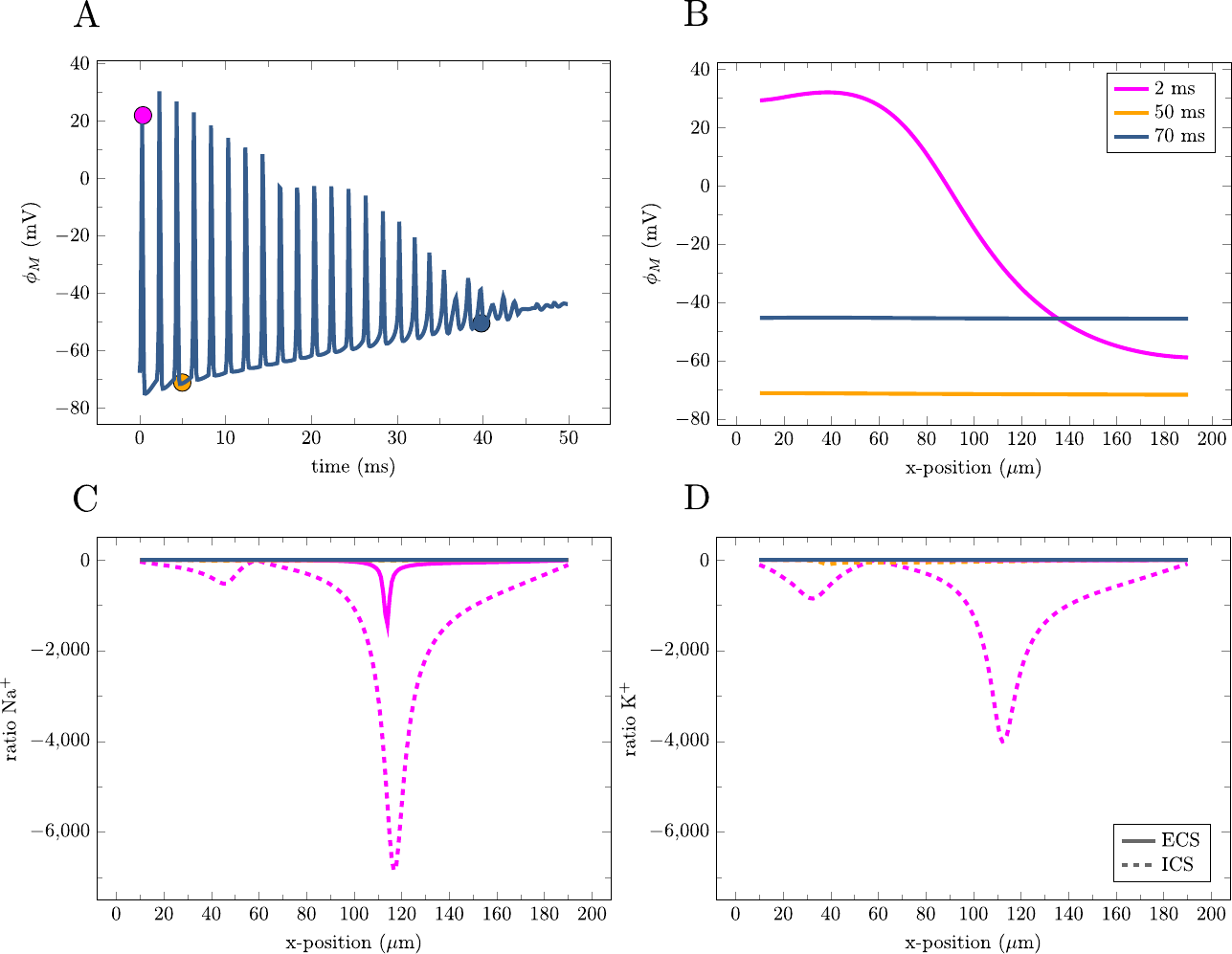}
    \label{fig:peclet}
    \vspace{-20pt}
    \caption{Diffusive and advective contributions of ion transport in model C (idealized 3D axons). The two upper panels display the temporal (\textbf{A}) and spatial (\textbf{B}) evolution of the membrane potential recorded at respectively the point $x=(100, 0.5, 0.5)$ $\mu$m and at 20 ms (pink), 50 ms (orange) and 70 ms (blue). The lower panels display the ratio between diffusion and advection for transport of $\rm Na^+$ (\textbf{C}) and $\rm K^+$ (\textbf{D}) where solid lines and dashed lines are respectively ECS and ICS traces of the ratio along the $x$-axis. Recall that the ratio is positive where diffusion dominates, and negative where drift dominates.}
\end{figure}

To assess the advective and diffusive contributions to ion transport in
physiologically relevant setting for excitable tissue, we run simulations using
model C and calculate the (spatially and temporally varying) ratio given
by~\eqref{eq:peclet:new} for three different time-points: (i) at 3 ms when an
action potential peaks, (ii) at 50 ms when the neuron is at rest, and (iii) at
70 ms where the neuron is severely depolarized. Recall that the ratio is
negative where advection dominates (cf.~\Cref{sec:methods:peclet}). We
observe that advection dominates locally in space during action potential
firing, for both ICS and ECS $\rm Na^{+}$,~$\rm
K^{+}$~(\Cref{fig:peclet}), and $\rm Cl^{-}$ (results not shown).
Specifically, the ECS ratio during an action potential (at 3 ms) peaks at
$-6500$ and $-4000$ for $\rm Na^{+}$ and $\rm K^{+}$, respectively. The ratio
is close to zero outside the spatial and temporal zone of the peak, indicating
that the advection dominance is local and related to the positioning of the
action potential. 

\subsection{Physiologically realistic 3D simulation}
\begin{figure}
    \centering
\includegraphics[width=0.98\linewidth]{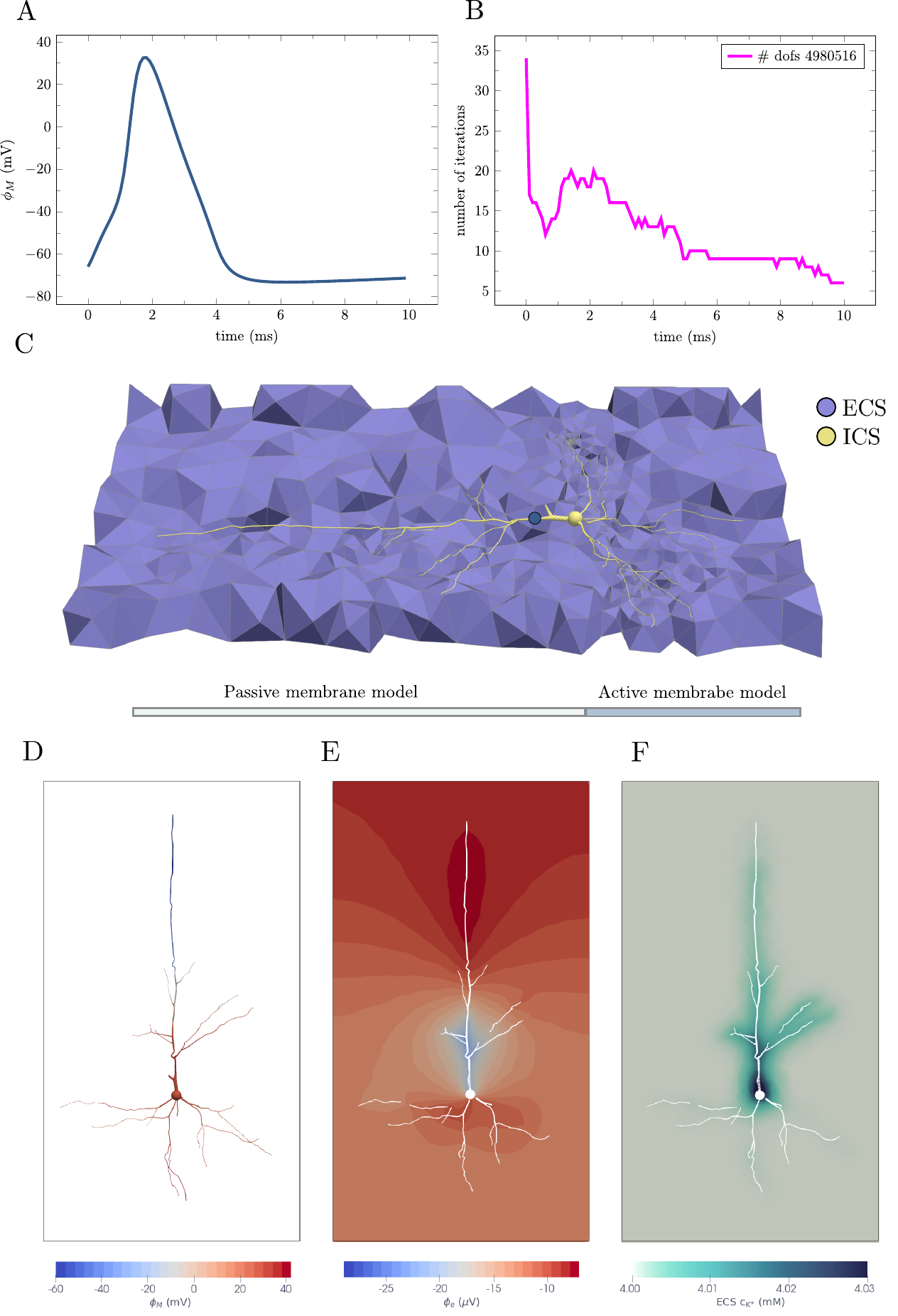}
\label{fig:numit:rat}
\vspace{-15pt}
    \caption{Simulation results with morphologically realistic 3D geometry (model D). The figure displays the temporal evolution of the membrane potential at point $(25, 0.3, 0.4)$ $\mu$m (\textbf{A}),
    the total number of iterations for the full KNP-EMI system over simulation time (\textbf{B}), an illustration of the problem domain and mesh (\textbf{C}), snapshots of the membrane potential (\textbf{D}) and the extracellular potential (\textbf{E})  at $t=1.7$ ms, and finally a snapshot of the ECS $K^+$ concentration  at $t=20$ ms (\textbf{F}).}
\end{figure}

We here consider a geometry based on a digital reconstruction of a pyramidal neuron (model D) to
assess how our proposed solver behaves on a realistic geometry. 
Action potentials are induced every $20$ ms near the tip of the dendrites. The resulting membrane depolarization spreads in space with a conduction velocity of $0.539$ m/s along the axon (\Cref{fig:numit:rat}A,D). The neuronal activity effects the ECS potential locally (\Cref{fig:numit:rat}E) and causes an increase in the ECS $\rm{K}^{+}$ concentration: after $20$ ms the concentration peaks at $4.03$ mM, notably increasing most near the axon and soma (\Cref{fig:numit:rat}F). 

Similarly to the models with idealized geometries (models B and C), we observe that the number of iterations before convergence in model D varies with the physical dynamics in the system: the number of iterations peaks at $20$ during the peak of the action potential (at $2$ ms) and decreases to $6$ when the neuron is at rest (\Cref{fig:numit:rat}B). 

\subsection{Properties of the DG scheme in realistic brain tissue geometries}

\begin{figure}
\includegraphics[width=1.0\linewidth]{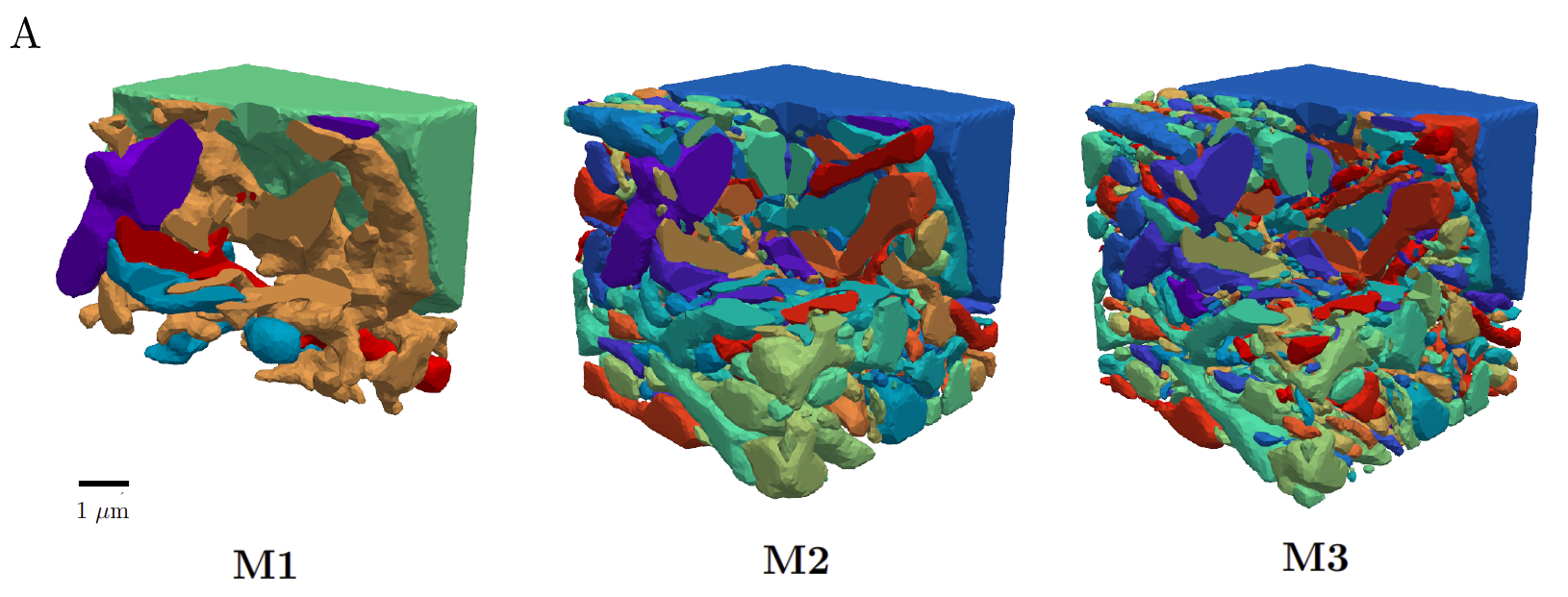} \\
   B
   \vspace{0.2cm}
    \begin{center}
    \begin{tabular}{lllllll}
    \toprule
    mesh & \# cells & \# dofs mortar & \# dofs !LM & \# dofs DG & r$_{\rm mortar}$ & r$_{\rm !LM}$  \\
    \midrule
    M1 &  165211 & 249076 & 223512 & 1982532  & 7.96 & 8.87\\
    M2 &  508644 & 768698 & 685896 &  6103728 & 7.95 & 8.90 \\
    M3 &  622294 & 876556 &835472& 7467528 & 8.52 & 8.94 \\
       Model C & 124416 & 108621 &  105668& 1492992 & 13.7& 14.1 \\
\bottomrule 
    \end{tabular}
    \caption{
      Comparison of system size for the mortar~\cite{ellingsrud2020finite}, multiplier-free~\cite{benedusi2024scalable} and the DG schemes in
      the case of interface dominated geometries. First order elements are used. Illustrations of interface dominated geometries with 5 (M1), 50 (M2), and 200 (M3) brain
      cells (ICS) surrounded by ECS (\textbf{A}). Comparison of the number of mesh cells ($\#$ cells), the number of degrees
      of freedom for the mortar scheme ($\#$ dofs mortar), the conforming multiplier-free scheme ($\#$ dofs !LM) and the DG scheme ($\#$ dofs DG) as well as the ratio between the system sizes of mortar and
      DG (r$_{\rm mortar}$) and of multiplier-free and DG (r$_{\rm !LM}$) for the three interface
      dominated meshes and the idealized 3D mesh from Model C (\textbf{B}).}
\label{tab:dof}                 
    \end{center}
\end{figure}

Brain tissue consist of tightly packed cells separated by thin sheets of ECS. Realistic geometries for numerical simulations of dynamics in collections of brain cells will as such typically be interface dominated (\Cref{tab:dof}A).
In $H^1$-conforming formulations of the KNP-EMI equations (e.g.~the mortar formulation~\cite{ellingsrud2020finite}
and the multiplier-free formulation~\cite{benedusi2024scalable}), the degrees of freedom at the membrane interfaces will be doubled as they count both in the intra- and
extracellular systems, see also \Cref{fig:domain_dofs}. Using linear polynomials, we next assess the cost (in terms of system size) of applying DG, which
typically leads to a greater number of degrees of freedom (dofs) than conforming formulations, for these
interface dominated geometries. We find that the DG discretization results in systems that are
$\sim 8$ times larger than that of the conforming formulations (\Cref{tab:dof}B).
The same factor is $\sim 14$ for the idealized $3$D mesh in model D (\Cref{tab:dof}B). The cost
(in terms of system size) of DG is thus reduced when considering interface dominated geometries. 

\section{Conclusions and outlook}


We have presented a novel solution strategy for solving the KNP-EMI equations which enables their large-scale
simulations. The key components of the strategy are: (i) a splitting scheme to decouple the NP equations governing ion
transport and the equation arising from the electro-neutrality assumption, (ii) a single dimensional/multiplier-free
formulation of the decoupled systems and their discontinuous Galerkin discretization, and (iii) their robust and scalable solvers. 
Numerical investigations show that the proposed discretization scheme is accurate and converges at expected rates in
relevant norms, and that the solution strategy is robust with respect to discretization parameters -- both for idealized and realistic two and three dimensional geometries. Strong and weak scaling experiments further show that the parallel scalability of the solver is close to optimal.

Previously the KNP-EMI problem has been solved monolithically~\cite{mori2009numerical, ellingsrud2020finite, benedusi2024EMI}. We here introduced a splitting scheme resulting in two smaller and importantly more standard sub-problems, namely an EMI problem and a series of advection diffusion problems, that are discretized and solved using established techniques. For relevant time steps in the neuroscience scenarios considered here, we observe that the splitting scheme is stable. As applications of KNP-EMI models extend beyond neuroscience, to e.g.~modelling of Li-ion batteries~\cite{wiedemann2013effects, kespe2019three}, where other temporal scales may apply, obtaining a better understanding of the accuracy and stability of the splitting scheme via rigorous analysis would be of interest. 

The DG scheme is flexible in the sense that its implementation only requires standard functionally in the
finite element software used. Further, the DG scheme ensures local mass conservation in each of the sub-problems. However,
the DG scheme results in systems with more degrees of freedom than conforming discretizations. A thorough comparison, addressing e.g.~accuracy, computational cost and conservation properties, of the previously presented schemes~\cite{ellingsrud2020finite, benedusi2024EMI, huynh2023convergence, de2024boundary} would be valuable.

Concerning practical applications, highly detailed reconstructions of brain tissue and cellular geometries have become available with recent advancements in image technology (see e.g.~\cite{microns2021functional, motta2019dense}). New and scalable solution algorithms for geometrically explicit models, such as the KNP-EMI model, allow for new high-fidelity~\emph{in-silico} models taking full advantage of data-sets describing the tissue morphology. Such models could be used to e.g.~study how the morphology of cellular processes or the spatial distribution of ionic membrane channels affect transport and buffering in brain tissue, potentially giving new insight into brain signalling and homeostasis in physiology and pathology. 

\section*{Acknowledgments} 
Ada J.~Ellingsrud and Pietro Benedusi acknowledges support from the Research
Council of Norway via FRIPRO grant \#324239 (EMIx) and from the national
infrastructure for computational science in Norway, Sigma2, via grant
\#NN8049K. Miroslav Kuchta acknowledges support from the Research Council of
Norway via DataSim \#303362.  We thank Rami Masri for valuable discussions and
input on discontinuous Galerkin methods and Jørgen Dokken for his contributions
in setting up the associated software repository.

\section*{Conflict of interest}
The authors declare that they have no conflict of interest.

\bibliographystyle{siamplain}
\bibliography{main}
\end{document}